\documentclass[aps,prd,onecolumn,groupedaddress,showpacs,nofootinbib,amssymb]{revtex4}
\usepackage[dvips]{graphicx}
\usepackage{amssymb}
\usepackage{amsmath}
\usepackage{graphicx,,color}
\usepackage{amsfonts}
\usepackage{bm}
\usepackage{cancel}
\usepackage{comment}

\newcommand\be{\begin{equation}}
\newcommand\ee{\end{equation}}

\allowdisplaybreaks[4]

\begin{document}

\title{Primordial Gravitational Waves Predictions for GW170817-compatible Einstein-Gauss-Bonnet Theory}
\author{V.K. Oikonomou,$^{1}$}\email{v.k.oikonomou1979@gmail.com,voikonomou@auth.gr}
\affiliation{$^{1)}$Department of Physics, Aristotle University of
Thessaloniki, Thessaloniki 54124, Greece}

\tolerance=5000

\begin{abstract}
In this work we shall calculate in detail the effect of an
GW170817-compatible Einstein-Gauss-Bonnet theory on the energy
spectrum of the primordial gravitational waves. The spectrum is
affected by two characteristics, the overall amplification/damping
factor caused by the GW170817-compatible Einstein-Gauss-Bonnet
theory and by the tensor spectral index and the tensor-to-scalar
ratio. We shall present the formalism for studying the
inflationary dynamics and post-inflationary dynamics of
GW170817-compatible Einstein-Gauss-Bonnet theories for all
redshifts starting from the radiation era up to the dark energy
era. We exemplify our formalism by using two characteristic
models, which produce viable inflationary and dark energy eras. As
we demonstrate, remarkably the overall damping/amplification
factor is of the order of unity, thus the GW170817-compatible
Einstein-Gauss-Bonnet models affect the primordial gravitational
waves energy spectrum only via their tensor spectral index and the
tensor-to-scalar ratio. Both models have a blue tilted tensor
spectrum, and therefore the predicted energy spectrum of the
primordial gravity waves can be detectable by most of the future
gravitational waves experiments, for various reheating
temperatures.
\end{abstract}

\pacs{04.50.Kd, 95.36.+x, 98.80.-k, 98.80.Cq,11.25.-w}

\maketitle

\section{Introduction}

Undoubtedly General Relativity (GR) is the most correct
description for astrophysical phenomena and objects, however even
at the astrophysical level, already appear examples that cast
doubt on its full validity at astrophysical levels, see for
example \cite{Abbott:2020khf}. Furthermore, GR fails to describe
in a consistent way persisting large scale phenomena in the
Universe, such as dark energy. The GR description of dark energy
is problematic, since if dark energy is phantom, phantom scalar
fields must be used to describe it, and phantom fields are not
appealing. On the other hand, if one sticks with GR descriptions
of the primordial era of the Universe, and specifically the
inflationary era
\cite{inflation1,inflation2,inflation3,inflation4}, inevitably the
scalar field description is the only option. The inflaton
description however is deemed problematic, due to the tremendous
fine tuning required for the inflaton interactions with other
particles, its mass and its very own self interaction scalar
potential. Although scalar fields frequently occur in string
theory, which is the most correct theoretical description of a
unified theory of everything, it seems that a simple GR
description might not be the complete answer for inflation.
Conceptually, single scalar fields are expected to exist in the
post-Planck era, as remnants of the UV-complete theory. However,
scalar fields are not the only constituents of the post-Planck
inflationary Lagrangian, since higher order curvature terms might
also be present in the effective inflationary Lagrangian. In fact,
these quantum terms are either higher powers of the Ricci scalar
$R$, square or cubic powers, and also Gauss-Bonnet terms. Thus,
modified gravity
\cite{reviews1,reviews2,reviews3,reviews4,reviews5,reviews6} might
eventually be the correct description of nature primordially in
the post-Planck era. The so-called Einstein-Gauss-Bonnet theories
thus offer another viable description of the inflationary era,
combining string motivated quantum corrections and scalar fields
\cite{Hwang:2005hb,Nojiri:2006je,Cognola:2006sp,Nojiri:2005vv,Nojiri:2005jg,Satoh:2007gn,Bamba:2014zoa,Yi:2018gse,Guo:2009uk,Guo:2010jr,Jiang:2013gza,Kanti:2015pda,vandeBruck:2017voa,Koh:2014bka,Bayarsaikhan:2020jww,Kanti:1998jd,Pozdeeva:2020apf,Vernov:2021hxo,Pozdeeva:2021iwc,Fomin:2020hfh,DeLaurentis:2015fea,Chervon:2019sey,Nozari:2017rta,Odintsov:2018zhw,Kawai:1998ab,Yi:2018dhl,vandeBruck:2016xvt,Kleihaus:2019rbg,Bakopoulos:2019tvc,Maeda:2011zn,Bakopoulos:2020dfg,Ai:2020peo,Oikonomou:2020oil,Odintsov:2020xji,Oikonomou:2020sij,Odintsov:2020zkl,Odintsov:2020mkz,Venikoudis:2021irr,Easther:1996yd,Antoniadis:1993jc,Antoniadis:1990uu,Kanti:1995vq,Kanti:1997br,Odintsov:2020sqy,Oikonomou:2021kql,Kong:2021qiu}.
However, Einstein-Gauss-Bonnet theories have a serious flaw
related with the propagation speed of the primordial tensor
perturbations, which is not equal to that of light's. The GW170817
event indicated that the propagation speed of gravitational waves
is almost identical to that of light's, therefore since there is
no fundamental particle physics reason for the graviton to change
its mass in the post-inflationary era, the graviton must be
massless primordially. Ordinary Einstein-Gauss-Bonnet theories
have this flaw related to the propagation speed of the tensor
modes. However, in our recent works
\cite{Odintsov:2020sqy,Oikonomou:2021kql} we provided a
theoretical remedy for the problem of massive primordial
gravitons. Specifically we showed that if the scalar potential and
scalar coupling function are related in a specific way, the speed
of the tensor perturbations can be equal to unity in natural
units, for Einstein-Gauss-Bonnet theories.

Now apart from the theoretical problems of the GR description of
inflation using solely scalar fields, there might be another
problem for it in the near future. The future gravitational wave
experiments
\cite{Hild:2010id,Baker:2019nia,Smith:2019wny,Crowder:2005nr,Smith:2016jqs,Seto:2001qf,Kawamura:2020pcg,Bull:2018lat}
will seek for stochastic gravitational waves in the Universe. If a
signal is detected by the future experiments, then single scalar
field descriptions will be in a very bad position, since the
current predictions of those theories lead to undetectable
signals. A possible signal will either mean that some modified
gravity is the underlying theory that produces the signal, or that
some abnormal reheating era takes place. Even if the latter
scenario occurs, the GR description produces a low scalar energy
spectrum of primordial gravitational waves. On the other hand,
theories such as the Einstein-Gauss-Bonnet theories lead to
detectable signals from future gravitational wave experiments.
Thus it is vital to thoroughly investigate all the aspects of
Einstein-Gauss-Bonnet theories.

In this work we shall consider the GW170817-compatible
Einstein-Gauss-Bonnet theories developed in Ref.
\cite{Oikonomou:2021kql}, and we shall calculate the predicted
energy spectrum of the primordial gravitational waves. We shall
consider inflationary aspects of several well chosen models and
also we shall focus on models which also produce a viable
late-time era. Overall we shall provide a unified description of
the early and late-time accelerations with the same models.
However the major issue we shall address in this work is that we
shall calculate numerically the overall amplification or damping
factor of the primordial gravitational wave signal, starting from
zero redshift and up to the radiation domination era. This is the
first time that such a calculation is performed for viable
Einstein-Gauss-Bonnet theories. In this way, we shall have a
complete picture for these inflationary theories.

This paper is organized as follows: In section II we shall present
the essential features of the inflationary dynamics for
GW170817-compatible Einstein-Gauss-Bonnet theories. In the same
section, the formalism for describing the post-inflationary and
late-time dynamics is described in detail. In section III two
characteristic Einstein-Gauss-Bonnet models are studied in detail,
and specifically their late-time and inflationary behavior is
studied. In section IV, the formalism for extracting the energy
spectrum of the primordial gravitational waves for
Einstein-Gauss-Bonnet is described in detail. We present the exact
formulas needed for the calculation of the overall
amplification/damping factor caused by the Einstein-Gauss-Bonnet
theory and we present the predicted energy spectrum of the
primordial gravitational waves for the models developed in section
IV. Finally the conclusions follow at the end of the paper.

\section{Essential Features of GW170817-compatible Einstein-Gauss-Bonnet Inflation and the Dark Energy Era}

\subsection{Inflationary Phenomenology Theoretical Framework of GW170817-compatible Einstein-Gauss-Bonnet Theory}

In this section we shall present the essential features of an
GW170817-compatible Einstein-Gauss-Bonnet gravity, which was
developed in \cite{Oikonomou:2021kql}, see also
\cite{Odintsov:2020sqy} for an alternative approach. First we
consider the Einstein-Gauss-Bonnet action,
\begin{equation}
\label{action} \centering
S=\int{d^4x\sqrt{-g}\left(\frac{R}{2\kappa^2}-\frac{1}{2}\partial_{\mu}\phi\partial^{\mu}\phi-V(\phi)-\frac{1}{2}\xi(\phi)\mathcal{G}\right)}\,
,
\end{equation}
with $R$ denoting as usual the Ricci scalar,
$\kappa=\frac{1}{M_p}$ with $M_p$ being the reduced Planck mass.
Also $\mathcal{G}$ denotes the Gauss-Bonnet invariant in four
dimensions, which is
$\mathcal{G}=R^2-4R_{\alpha\beta}R^{\alpha\beta}+R_{\alpha\beta\gamma\delta}R^{\alpha\beta\gamma\delta}$
with $R_{\alpha\beta}$ and $R_{\alpha\beta\gamma\delta}$ denoting
the Ricci and Riemann tensor. For the whole analysis that follows
we shall assume that the geometric background is described by a
flat Friedmann-Robertson-Walker (FRW) metric, with line element,
\begin{equation}
\label{metric} \centering
ds^2=-dt^2+a(t)^2\sum_{i=1}^{3}{(dx^{i})^2}\, ,
\end{equation}
where $a(t)$ is the scale factor. Assuming that the scalar field
is solely time-dependent and by varying the gravitational action
with respect to the metric tensor and the scalar field, we obtain
the following equations of motion,
\begin{equation}
\label{motion1} \centering
\frac{3H^2}{\kappa^2}=\frac{1}{2}\dot\phi^2+V+12 \dot\xi H^3\, ,
\end{equation}
\begin{equation}
\label{motion2} \centering \frac{2\dot
H}{\kappa^2}=-\dot\phi^2+4\ddot\xi H^2+8\dot\xi H\dot H-4\dot\xi
H^3\, ,
\end{equation}
\begin{equation}
\label{motion3} \centering \ddot\phi+3H\dot\phi+V'+12 \xi'H^2(\dot
H+H^2)=0\, .
\end{equation}
For the whole analysis of the inflationary era, we shall assume
that the slow-roll conditions are valid,
\begin{equation}\label{slowrollhubble}
\dot{H}\ll H^2,\,\,\ \frac{\dot\phi^2}{2} \ll V,\,\,\,\ddot\phi\ll
3 H\dot\phi\, .
\end{equation}
Moreover, the speed of the primordial tensor modes for the
Einstein-Gauss-Bonnet theory has the following functional form,
\begin{equation}
\label{GW} \centering c_T^2=1-\frac{Q_f}{2Q_t}\, ,
\end{equation}
where $Q_f=8 (\ddot\xi-H\dot\xi)$, $Q_t=F+\frac{Q_b}{2}$,
$F=\frac{1}{\kappa^2}$ and $Q_b=-8 \dot\xi H$. In order for the
Einstein-Gauss-Bonnet theory to be compatible with the GW170817
event, we must require $c_T^2=1$ in natural units, therefore this
can be achieved if $Q_f=0$ which in turn results to the following
differential equation $\ddot\xi=H\dot\xi$. Clearly this
differential equation constrains the scalar coupling function $\xi
(\phi)$. Expressed in terms of the scalar field, the constraint
equation reads,
\begin{equation}
\label{constraint1} \centering
\xi''\dot\phi^2+\xi'\ddot\phi=H\xi'\dot\phi\, ,
\end{equation}
where the ``prime'' indicates differentiation with respect to the
scalar field. Motivated the the slow-roll conditions of the scalar
field, by assuming that,
\begin{equation}\label{firstslowroll}
 \xi'\ddot\phi \ll\xi''\dot\phi^2\, ,
\end{equation}
the constraint (\ref{constraint1}) is rewritten as,
\begin{equation}
\label{constraint} \centering
\dot{\phi}\simeq\frac{H\xi'}{\xi''}\, ,
\end{equation}
hence by combining Eqs. (\ref{motion3}) and (\ref{constraint}) we
get,
\begin{equation}
\label{motion4} \centering \frac{\xi'}{\xi''}\simeq-\frac{1}{3
H^2}\left(V'+12 \xi'H^4\right)\, .
\end{equation}
Furthermore we shall assume that the following conditions hold
true for the theory at hand,
\begin{equation}\label{mainnewassumption}
\kappa \frac{\xi '}{\xi''}\ll 1\, ,
\end{equation}
\begin{equation}\label{scalarfieldslowrollextra}
12 \dot\xi H^3=12 \frac{\xi'^2H^4}{\xi''}\ll V\, .
\end{equation}
In view of Eqs. (\ref{slowrollhubble}), (\ref{constraint}) and
(\ref{scalarfieldslowrollextra}), the equations of motion take the
following form,
\begin{equation}
\label{motion5} \centering H^2\simeq\frac{\kappa^2V}{3}\, ,
\end{equation}
\begin{equation}
\label{motion6} \centering \dot H\simeq-\frac{1}{2}\kappa^2
\dot\phi^2\, ,
\end{equation}
\begin{equation}
\label{motion8} \centering \dot\phi\simeq\frac{H\xi'}{\xi''}\, ,
\end{equation}
while the condition (\ref{scalarfieldslowrollextra}) becomes,
\begin{equation}\label{mainconstraint2}
 \frac{4\kappa^4\xi'^2V}{3\xi''}\ll 1\, .
\end{equation}
Also the constraint differential equation (\ref{motion4}) which
relates the functional forms of the scalar coupling function and
of the scalar potential, becomes,
\begin{equation}
\label{maindiffeqnnew} \centering
\frac{V'}{V^2}+\frac{4\kappa^4}{3}\xi'\simeq 0\, .
\end{equation}
In view of the above simplifications, the slow-roll indices for
the Einstein-Gauss-Bonnet inflationary framework at hand take the
following simplified form \cite{Oikonomou:2021kql},
\begin{equation}
\label{index1} \centering \epsilon_1\simeq\frac{\kappa^2
}{2}\left(\frac{\xi'}{\xi''}\right)^2\, ,
\end{equation}
\begin{equation}
\label{index2} \centering
\epsilon_2\simeq1-\epsilon_1-\frac{\xi'\xi'''}{\xi''^2}\, ,
\end{equation}
\begin{equation}
\label{index3} \centering \epsilon_3=0\, ,
\end{equation}
\begin{equation}
\label{index4} \centering
\epsilon_4\simeq\frac{\xi'}{2\xi''}\frac{\mathcal{E}'}{\mathcal{E}}\,
,
\end{equation}
\begin{equation}
\label{index5} \centering
\epsilon_5\simeq-\frac{\epsilon_1}{\lambda}\, ,
\end{equation}
\begin{equation}
\label{index6} \centering \epsilon_6\simeq
\epsilon_5(1-\epsilon_1)\, ,
\end{equation}
where $\mathcal{E}=\mathcal{E}(\phi)$ and $\lambda=\lambda(\phi)$
are defined as follows,
\begin{equation}\label{functionE}
\mathcal{E}(\phi)=\frac{1}{\kappa^2}\left(
1+72\frac{\epsilon_1^2}{\lambda^2} \right),\,\, \,
\lambda(\phi)=\frac{3}{4\xi''\kappa^2 V}\, .
\end{equation}
Having the slow-roll indices available, one can easily calculate
the observational indices of inflation. We shall mainly be
interested in the spectral index of the primordial scalar
curvature perturbations $n_{\mathcal{S}}$, the spectral index of
the tensor perturbations $n_{\mathcal{T}}$ and the
tensor-to-scalar ratio $r$, which in terms of the slow-roll
indices are expressed as follows,
\begin{equation}
\label{spectralindex} \centering
n_{\mathcal{S}}=1-4\epsilon_1-2\epsilon_2-2\epsilon_4\, ,
\end{equation}
\begin{equation}\label{tensorspectralindex}
n_{\mathcal{T}}=-2\left( \epsilon_1+\epsilon_6 \right)\, ,
\end{equation}
\begin{equation}\label{tensortoscalar}
r=16\left|\left(\frac{\kappa^2Q_e}{4H}-\epsilon_1\right)\frac{2c_A^3}{2+\kappa^2Q_b}\right|\,
,
\end{equation}
with $c_A$ being the sound speed,
\begin{equation}
\label{sound} \centering c_A^2=1+\frac{Q_aQ_e}{3Q_a^2+
\dot\phi^2(\frac{2}{\kappa^2}+Q_b)}\, ,
\end{equation}
with,
\begin{align}\label{qis}
& Q_a=-4 \dot\xi H^2,\,\,\,Q_b=-8 \dot\xi H,\,\,\,
Q_t=F+\frac{Q_b}{2},\\
\notag &  Q_c=0,\,\,\,Q_e=-16 \dot{\xi}\dot{H}\, .
\end{align}
As it was shown in Ref. \cite{Oikonomou:2021kql}, by using the
previous simplifications, the tensor-to-scalar ratio takes the
following simplified form,
\begin{equation}\label{tensortoscalarratiofinal}
r\simeq 16\epsilon_1\, ,
\end{equation}
while the spectral index of the primordial tensor perturbations
takes the following simplified form,
\begin{equation}\label{tensorspectralindexfinal}
n_{\mathcal{T}}\simeq -2\epsilon_1\left ( 1-\frac{1}{\lambda
}+\frac{\epsilon_1}{\lambda}\right)\, ,
\end{equation}
with $\lambda$ being defined in Eq. (\ref{functionE}). Also, the
$e$-foldings number can be evaluated in terms of the scalar
coupling function $\xi (\phi)$ as follows,
\begin{equation}
\label{efolds} \centering
N=\int_{t_i}^{t_f}{Hdt}=\int_{\phi_i}^{\phi_f}{\frac{\xi''}{\xi'}d\phi}\,
,
\end{equation}
where $\phi_f$ and $\phi_i$ are the values of the scalar field at
the end of inflation and at the first horizon crossing during the
beginning of inflation, respectively. Having the theoretical
framework we just presented at hand, one can easily investigate if
a specific Einstein-Gauss-Bonnet model can be viable when compared
to the Planck 2018 data.

Before closing this section, an important remark is in order.
Specifically, in this work we considered mainly linear effects
governing the dynamics of Einstein-Gauss-Bonnet inflation and the
corresponding gravitational wave speed. So the constraint equation
(\ref{constraint1}) is valid only at a linear level, and indeed
this linear level would persist for gravitational waves with
wavelengths larger than $10\,$Mpc. However, for these modes which
are mainly probed by CMB experiments and are basically CMB modes,
the linear theory applies. Below $10\,$Mpc, one expects non-linear
effects to take place, in the form of higher order operators and
quantum corrections, see the comprehensive analysis in Refs.
\cite{Creminelli:2017sry,Ezquiaga:2017ekz}. Indeed, for
wavelengths corresponding to gravitational waves corresponding to
the GW170817 event, one should certainly include in the study of
cosmological gravity waves, the non-linear effects imposed by
higher order corrections and quantum corrections. The stability of
the solutions and the effects of these corrections on the
background evolution should be addressed carefully, taking also
into account the quantum and higher order corrections. Hence the
constraint (\ref{constraint1}) is explicitly background dependent
and only works up to the linear level. Therefore, any small
deviation coming from any non-linear effects will effectively
spoil this constraint relation. Hence, although the GW170817
measurement is particularly important since it directly probes the
gravitational wave speed over cosmological distances, one should
take into account that the gravitational wave speed might be
locally reduced in high energy environments. Also, although beyond
$10\,$Mpcs, screening effects are believed not to persist and
affect the gravitational wave speed, for low energy measurements
at distances 10 000 km, the effective field theory of dark energy
or modified gravity must be used in a compelling way
\cite{Creminelli:2017sry}. Hence the higher order corrections and
quantum corrections terms originating from the underlying
effective field theory of any origin, must be used in a formal
way. In this work though we need to clarify that these effects are
not taken into account. However, such an extension of the present
paper to include the effects of these operators is compelling,
since all the relevant modes to future gravitational waves
experiments have wavelengths smaller  that $ 10\,$Mpc, thus all
the quantum and higher order effects apply. We aim to formally
address this important issue in a future work.


\subsection{From Reheating to Dark Energy Phenomenology of GW170817-compatible Einstein-Gauss-Bonnet Theory}

In the previous subsection we presented the inflationary
theoretical framework for a general GW170817-compatible
Einstein-Gauss-Bonnet gravity. A crucial assumption for this
theory was that the tensor perturbations propagate with a speed
equal to that of light's. Since there is no fundamental reason for
the gravitons to propagate differently compared to light during
the post-inflationary era, in this section we shall develop a
formalism for studying the dark energy  and the previous eras up
to the inflationary era, for GW170817-compatible
Einstein-Gauss-Bonnet theories. The reason for studying the
post-inflationary eras up to the dark energy era, is mainly the
fact that in a later section we will need to quantify the effect
of the Einstein-Gauss-Bonnet gravity on the energy spectrum of the
primordial gravitational waves from reheating up to the dark
energy era.

The condition $\ddot\xi=H\dot\xi$ which ensures that primordially
$c_T^2=1$, reads,
\begin{equation}
\centering \label{dotxi} \dot\xi=\mathcal{C} e^{\int{Hdt}}\, ,
\end{equation}
with $\mathcal{C}$ being an integration constant. Since,
$\frac{dz}{dt}=-H(1+z)$, the above equation yields,
\begin{equation}
\centering \label{dotxi2}
\dot\xi=a(t)\mathcal{C}=\frac{\mathcal{C}}{1+z}\, .
\end{equation}
The above relation greatly simplifies the study of the dark energy
era, since it specifies the functional form of $\dot{\xi}$ which
enters the gravitational equations of motion. In view of Eq.
(\ref{dotxi2}), the field equations read,
\begin{equation}
\centering \label{motion8}
\frac{3H^2}{\kappa^2}=\rho_m+\frac{1}{2}\dot\phi^2+V+24\frac{\mathcal{C}}{1+z}H^3\,
,
\end{equation}
\begin{equation}
\centering \label{motion9} -\frac{2\dot
H}{\kappa^2}=\rho_m+P_m+\dot\phi^2-16\frac{\mathcal{C}}{1+z}H\dot
H\, ,
\end{equation}
\begin{equation}
\centering \label{motion10}
V_\phi+\ddot\phi+3H\dot\phi+\frac{\mathcal{C}}{1+z}\frac{\mathcal{G}}{\dot\phi}=0\,
.
\end{equation}
The parameter $\mathcal{C}$ is a free parameter of the theory, and
it proves that when it takes reasonable values, not fine-tuned to
be larger than $\sim \mathcal{O}(10^6)$, it does not crucially
affect the dark energy era. In order to study the
post-inflationary evolution of the Einstein-Gauss-Bonnet theory,
we shall express the field equations with respect to the redshift.
Hence we shall use the following,
\begin{equation}
\centering \dot H=-H(1+z)H'\, ,
\end{equation}
\begin{equation}
\centering \dot\phi=-H(1+z)\phi'\, ,
\end{equation}
\begin{equation}
\centering
\ddot\phi=H^2(1+z)^2\phi''+H^2(1+z)\phi'+HH'(1+z)^2\phi'\, ,
\end{equation}
\begin{equation}
\centering \dot
R=6H(1+z)^2\left(HH''+(H')^2-\frac{3HH'}{1+z}\right)\, ,
\end{equation}
in order to transform the field equations in terms of the
redshift. Also we shall introduce the statefinder quantity
$y_H(z)$ \cite{Bamba:2012qi,Odintsov:2020qyw},
\begin{equation}
\centering \label{yH} y_H=\frac{\rho_{DE}}{\rho_{d0}}\, ,
\end{equation}
where $\rho_{DE}$ denotes the energy density of dark energy, which
is in the case at hand,
\begin{equation}
\centering \label{DEdensity}
\rho_{DE}=\frac{1}{2}\dot\phi^2+V+24\dot\xi H^3\, ,
\end{equation}
while $\rho_{d0}$ is the value of density for non-relativistic
matter at present day. The pressure for the dark energy fluid is,
\begin{equation}
\centering \label{PDE} P_{DE}=-V-24\dot\xi H^3-8\dot\xi H\dot H\,
,
\end{equation}
with the dark energy fluid satisfying,
\begin{equation}
\centering \label{conteqDE} \dot\rho_{DE}+3H(\rho_{DE}+P_{DE})=0\,
.
\end{equation}
Therefore, equations (\ref{motion1}) and (\ref{motion2}) can be
written in the Friedmann equation-like form of standard
Einstein-Hilbert gravity,
\begin{equation}
\centering \label{motion6}
\frac{3H^2}{\kappa^2}=\rho_{(m)}+\rho_{DE}\, ,
\end{equation}
\begin{equation}
\centering \label{motion7} -\frac{2\dot
H}{\kappa^2}=\rho_{(m)}+P_{(m)}+\rho_{DE}+P_{DE}\, ,
\end{equation}
The Hubble rate and its derivatives entering in the field
equations, can be expressed in terms of the statefinder quantity
$y_H(z)$ as follows,
\begin{equation}
\centering \label{H}
H^2=m_s^2\left(y_H(z)+\frac{\rho_{(m)}}{\rho_{d0}}\right)\, ,
\end{equation}
\begin{equation}
\centering \label{H'}
HH'=\frac{m_s^2}{2}\left(y_H'+\frac{\rho_{(m)}'}{\rho_{d0}}\right)\,
,
\end{equation}
\begin{equation}
\centering \label{H''}
H'^2+HH''=\frac{m_s^2}{2}\left(y_H''+\frac{\rho_{(m)}''}{\rho_{d0}}\right)\,
,
\end{equation}
with $m_s^2=\kappa^2\frac{\rho_{d0}}{3}=1.87101\cdot10^{-67}$. In
the next section we shall numerically solve the differential
equations (\ref{motion4}) and (\ref{motion5}) with respect to the
scalar field and the statefinder quantity $y_H$, for the redshift
range from zero up to several millions. In this way, we will find
the behavior of the model from the dark energy era up to the
reheating era. For our study we shall consider several important
quantities characterizing the geometrical contribution of the
Einstein-Gauss-Bonnet gravity to the post-inflationary evolution
and to the dark energy era. We shall consider the equation of
state parameter $\omega_{DE}$ and the dark energy density
parameter $\Omega_{DE}$, which expressed as functions of the
statefinder quantity $y_H(z)$ are defined as follows
\cite{Bamba:2012qi,Odintsov:2020qyw},
\begin{align}
\centering \label{DE}
\omega_{DE}&=-1+\frac{1+z}{3}\frac{d\ln{y_H}}{dz}&\Omega_{DE}&=\frac{y_H}{y_H+\frac{\rho_{(m)}}{\rho_{d0}}}\,
.
\end{align}
Also for the late-time study, we shall consider another important
statefinder quantity, the deceleration parameter, defined as,
\begin{align}
\centering q&=-1-\frac{\dot H}{H^2}\, .
\end{align}
In the next sections we shall study the inflationary and
post-inflationary evolution of two Einstein-Gauss-Bonnet models
and we shall set the stage for the study of the predicted energy
spectrum of the primordial gravitational waves for these models.

\section{Study of the Phenomenology for Specific Einstein-Gauss-Bonnet Models: From Inflation to Dark Energy}

In this section we shall consider the evolution and phenomenology
of two distinct Einstein-Gauss-Bonnet models, starting from
inflation until the dark energy era. For the inflationary era, we
shall be interested in calculating the spectral index of the
primordial scalar curvature perturbations and the tensor-to-scalar
ratio of both models, which will be essential for the calculation
of the primordial gravitational wave energy spectrum. Recall that
the Planck 2018 collaboration \cite{Planck:2018jri} constrains the
spectral index and the tensor-to-scalar ratio as follows,
\begin{equation}\label{planck2018}
\centering n_{\mathcal{S}}=0.9649\pm0.0042,\,\,\, r<0.064\, .
\end{equation}
Accordingly, we shall numerically solve the equations of motion
for the post-inflationary era, which will enable us to measure the
effect of the Einstein-Gauss-Bonnet theory on the energy spectrum
of the primordial gravitational waves.

\subsection{Model I}

\begin{figure}
\centering
\includegraphics[width=18pc]{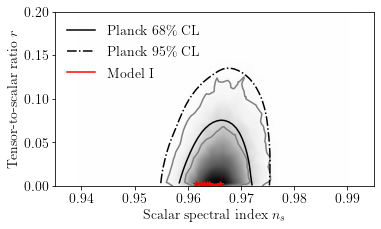}
\caption{Marginalized curves of the Planck 2018 data for the Model
I (red curve).}\label{plotplanck1}
\end{figure}

\begin{figure}[h!]
\centering
\includegraphics[width=20pc]{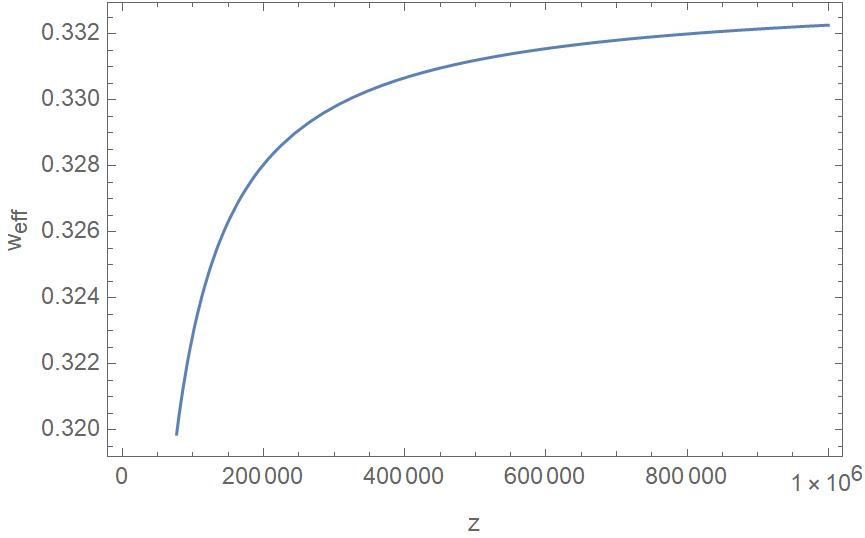}
\includegraphics[width=20pc]{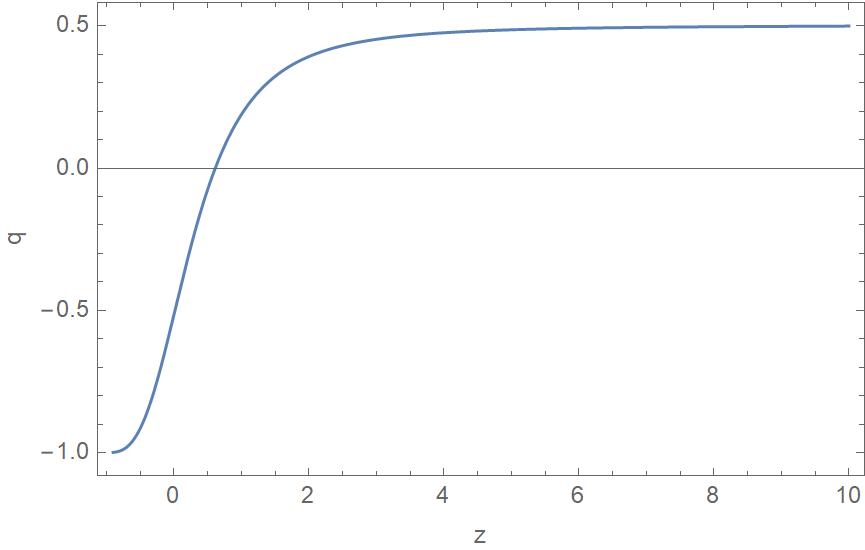}
\caption{The total EoS parameter $w_{eff}$ (left plot) and the
deceleration parameter $q$ (right plot), as functions of the
redshift.} \label{plot2}
\end{figure}

Let us start with an exponential functional form for the scalar
coupling function $\xi(\phi)$, which has the following form,
\begin{equation}
\label{modelA} \xi(\phi)=\beta  \exp \left(\left(\frac{\phi
}{M}\right)^2\right)\, ,
\end{equation}
with $\beta$ being a dimensionless parameter, and $M$ is a
parameter with mass dimensions $[m]^1$. Combining (\ref{modelA})
and Eq. (\ref{maindiffeqnnew}) and by solving the resulting
differential equation, the scalar potential is obtained, which is,
\begin{equation}
\label{potA} \centering V(\phi)=\frac{3}{3 \gamma  \kappa ^4+4
\beta  \kappa ^4 e^{\frac{\phi ^2}{M^2}}} \, ,
\end{equation}
with $\gamma$ being a dimensionless integration constant.
Accordingly, the slow-roll indices (\ref{index1})-(\ref{index6})
become,
\begin{equation}
\label{index1A} \centering \epsilon_1\simeq \frac{\kappa ^2 M^4
\phi ^2}{2 \left(M^2+2 \phi ^2\right)^2} \, ,
\end{equation}
\begin{equation}
\label{index2A} \centering \epsilon_2\simeq \frac{M^4
\left(2-\kappa ^2 \phi ^2\right)-4 M^2 \phi ^2}{2 \left(M^2+2 \phi
^2\right)^2}\, ,
\end{equation}
\begin{equation}
\label{index3A} \centering \epsilon_3=0\, ,
\end{equation}
\begin{equation}
\label{index5A} \centering \epsilon_5\simeq -\frac{4 \beta  \phi
^2 e^{\frac{\phi ^2}{M^2}}}{\left(M^2+2 \phi ^2\right) \left(3
\gamma +4 \beta  e^{\frac{\phi ^2}{M^2}}\right)} \, ,
\end{equation}
\begin{equation}
\label{index6A} \centering \epsilon_6\simeq -\frac{2 \beta  \phi
^2 e^{\frac{\phi ^2}{M^2}} \left(M^4 \left(2-\kappa ^2 \phi
^2\right)+8 M^2 \phi ^2+8 \phi ^4\right)}{\left(M^2+2 \phi
^2\right)^3 \left(3 \gamma +4 \beta  e^{\frac{\phi
^2}{M^2}}\right)} \, ,
\end{equation}
and we omitted the slow-roll index $\epsilon_4$ the functional
form of which was quite lengthy. Accordingly, by solving the
equation $\epsilon_1\simeq \mathcal{O}(1)$ yields the value of the
scalar field when inflation ends $\phi_f\simeq \frac{1}{4}
\sqrt{\kappa ^2 M^4+\kappa  M^3 \sqrt{\kappa ^2 M^2-16}-8 M^2}$.
Moreover, in order to calculate the value of the scalar field at
first horizon crossing, we shall use the $e$-foldings number Eq.
(\ref{efolds}), and by solving the resulting algebraic equation
with respect to $\phi_i$ we get $\phi_i=\frac{1}{4} M \sqrt{\kappa
^2 M^2+\kappa  M \sqrt{\kappa ^2 M^2-16}-16 Y-8}$. Moreover, the
scalar spectral index as a function of the scalar field reads,
\begin{align}\label{spectralpowerlawmodel}
& n_{\mathcal{S}}\simeq -1-\frac{\kappa ^2 M^4 \phi
^2}{\left(M^2+2 \phi ^2\right)^2}+\frac{4 \phi ^2 \left(3 M^2+2
\phi ^2\right)}{\left(M^2+2 \phi ^2\right)^2}\\ & \notag
+\frac{4608 \beta ^2 \phi ^6 e^{\frac{2 \phi ^2}{M^2}} \left(6
\gamma  \phi ^2+16 \beta  e^{\frac{\phi ^2}{M^2}} \left(M^2+\phi
^2\right)+9 \gamma  M^2\right)}{\left(M^2+2 \phi ^2\right)^4
\left(3 \gamma +4 \beta  e^{\frac{\phi ^2}{M^2}}\right)^3} \, ,
\end{align}
while the tensor spectral index reads,
\begin{align}\label{tensorspectralindexpowerlawmodel}
& n_{\mathcal{T}}\simeq \frac{\phi ^2 \left(-4 \beta e^{\frac{\phi
^2}{M^2}} \left(M^4 \left(3 \kappa ^2 \phi ^2-2\right)+\kappa ^2
M^6-8 M^2 \phi ^2-8 \phi ^4\right)-3 \gamma \kappa ^2 M^4
\left(M^2+2 \phi ^2\right)\right)}{\left(M^2+2 \phi ^2\right)^3
\left(3 \gamma +4 \beta  e^{\frac{\phi ^2}{M^2}}\right)}
 \, .
\end{align}
and the tensor-to-scalar ratio is,
\begin{equation}\label{tensortoscalarfinalmodelpowerlaw}
r\simeq \frac{8 \kappa ^2 M^4 \phi ^2}{\left(M^2+2 \phi
^2\right)^2}\, .
\end{equation}
The observational indices must be evaluated at the first horizon
crossing, so for $\phi=\phi_i$ and also the parameter $M$ is
assumed to be of the form $M=\mu/\kappa$ where $\mu$ is a
dimensionless free parameter. The model at hand is compatible with
the Planck data for a wide range of the free parameters. In Fig.
\ref{plotplanck1} we present the marginalized curves of Planck
2018 and the predictions of Model I, for the free parameters
taking values $\mu=[22.09147657871,22.09147657877]$, $\beta=-1.5$,
$\gamma=2$, for $N=60$ $e$-foldings. As it can be seen in Fig.
\ref{plotplanck1}, the model is fitted quite well within the
Planck 2018 data. Also for the above range of values for the free
parameters, the tensor spectral index takes values in the range
$n_{\mathcal{T}}=[0.378856,0.379088]$, hence it is positive. This
is of particular importance and shall play an important role for
energy spectrum of the primordial gravitational waves.
Let us now consider the evolution of Model I during the
post-inflationary era up to present day. For the calculation of
the impact of the Einstein-Gauss-Bonnet theory on the energy
spectrum of the primordial gravitational waves it is essential to
study the evolution of the model from present day, hence for $z=0$
up to high redshifts which stretch up to the reheating era, hence
for redshifts $z\sim 10^6$ or higher. For the numerical analysis
of the differential equations, we shall impose the following
initial conditions, $y_H(z=z_{fin})=\frac{\Lambda}{3m_s^2}$,
$\frac{dy_H}{dz}\Big|_{z_{fin}}=0$, $\phi(z=z_{fin})=10^{1.9}M_p$,
$\frac{d\phi}{dz}\Big|_{z=z_{fin}}=-10^{-6}M_p$, where
$z_{fin}=10^6$. The results of our numerical analysis for Model I
can be found in Fig. \ref{plot2} where we plot the total equation
of state (EoS) parameter $w_{eff}$ (left plot) and the
deceleration parameter (right plot) as functions of the redshift
$z$.
As it can be seen in Fig. \ref{plot2} the total EoS parameter
approaches the value $w_{eff}\sim 1/3$, hence the Model I for
redshifts $z\sim 10^6$ has entered deeply in the radiation
domination era. Also the Model I mimics the
$\Lambda$-Cold-Dark-Matter ($\Lambda$CDM) model at late times, as
it can be seen by looking the behavior of the deceleration
parameter. In Table \ref{Table1} we have gathered the values of
the dark energy EoS parameter $\omega_{DE}$, the dark energy
density parameter $\Omega_{DE}$ and of the deceleration parameter
$q$ at present day. Overall, Model I offers a viable description
regarding it's dark energy predictions.
\begin{table}[h!]
\caption{Model I Parameters}
\begin{center}
\begin{tabular}{|r|r|r|}
\hline \textbf{Parameter}&\textbf{Model I}&\textbf{$\Lambda$CDM Value or Planck 2018 Constraints}\\
\hline q(z=0)&-0.51626&-0.535
\\ \hline
 $\Omega_{DE}(0)$&0.67754&0.6847$\pm$0.0073\\ \hline
$\omega_{DE}(0)$&-1&-1.018$\pm$0.031\\ \hline
\end{tabular}
\label{Table1}
\end{center}
\end{table}

\subsection{Model II}

\begin{figure}
\centering
\includegraphics[width=18pc]{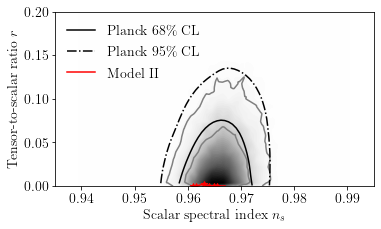}
\caption{Marginalized curves of the Planck 2018 data for the Model
II (red curve).}\label{plotplanck1power}
\end{figure}

\begin{figure}[h!]
\centering
\includegraphics[width=20pc]{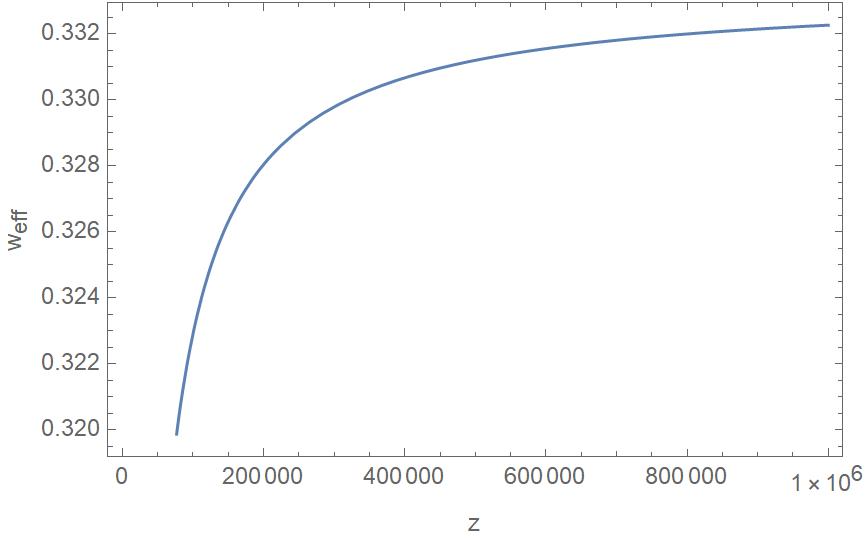}
\includegraphics[width=20pc]{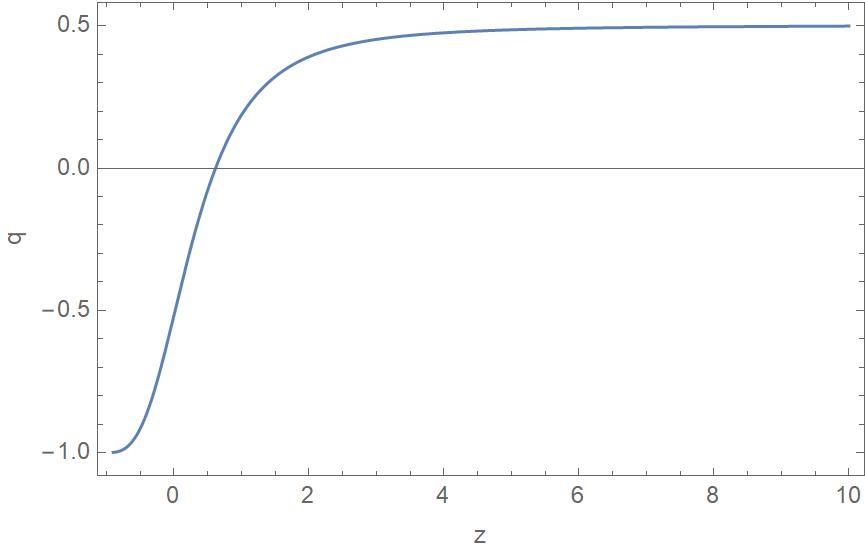}
\caption{The total EoS parameter $w_{eff}$ (left plot) and the
deceleration parameter $q$ (right plot), as functions of the
redshift.} \label{plot2power}
\end{figure}

Let us now consider a power law functional form for the scalar
coupling function $\xi(\phi)$, which has the following form,
\begin{equation}
\label{modelApower} \xi(\phi)=\beta  \left(\frac{\phi
}{M}\right)^{\nu }\, ,
\end{equation}
where $\beta$ is a dimensionless parameter, and $M$ is a parameter
with mass dimensions $[m]^1$. Upon combining (\ref{modelApower})
and Eq. (\ref{maindiffeqnnew}) and by solving the resulting
differential equation, the scalar potential is obtained, which in
this case is,
\begin{equation}
\label{potApower} \centering V(\phi)=\frac{3}{3 \gamma  \kappa
^4+4 \beta  \kappa ^4 \left(\frac{\phi }{M}\right)^{\nu }} \, ,
\end{equation}
with $\gamma$ being a dimensionless integration constant.
Accordingly, the slow-roll indices (\ref{index1})-(\ref{index6})
become in this case,
\begin{equation}
\label{index1Apower} \centering \epsilon_1\simeq \frac{\kappa ^2
\phi ^2}{2 (\nu -1)^2} \, ,
\end{equation}
\begin{equation}
\label{index2Apower} \centering \epsilon_2\simeq -\frac{\kappa ^2
\phi ^2-2 \nu +2}{2 (\nu -1)^2}\, ,
\end{equation}
\begin{equation}
\label{index3Apower} \centering \epsilon_3=0\, ,
\end{equation}
\begin{equation}
\label{index5Apower} \centering \epsilon_5\simeq-\frac{2 \beta
\nu  \left(\frac{\phi }{M}\right)^{\nu }}{(\nu -1) \left(3 \gamma
+4 \beta  \left(\frac{\phi }{M}\right)^{\nu }\right)} \, ,
\end{equation}
\begin{equation}
\label{index6Apower} \centering \epsilon_6\simeq -\frac{\beta  \nu
\left(-\kappa ^2 \phi ^2+2 \nu ^2-4 \nu +2\right) \left(\frac{\phi
}{M}\right)^{\nu }}{(\nu -1)^3 \left(3 \gamma +4 \beta
\left(\frac{\phi }{M}\right)^{\nu }\right)} \, ,
\end{equation}
and we omitted the slow-roll index $\epsilon_4$ in this model too,
because its functional form is quite lengthy. Accordingly, by
solving the equation $\epsilon_1\simeq \mathcal{O}(1)$ yields the
value of the scalar field when inflation ends $\phi_f\simeq
\frac{\sqrt{2} \sqrt{\nu ^2-2 \nu +1}}{\kappa }$. Also by using
the $e$-foldings number Eq. (\ref{efolds}), and by solving the
resulting algebraic equation with respect to $\phi_i$ we get
$\phi_i=\frac{\sqrt{2} \sqrt{(\nu -1)^2} e^{-\frac{Y}{\nu
-1}}}{\kappa }$. Moreover, the scalar spectral index as a function
of the scalar field reads,
\begin{align}\label{spectralpowerlawmodelpower}
& n_{\mathcal{S}}\simeq -\frac{\kappa ^2 \phi ^2}{(\nu
-1)^2}+\frac{2 (\nu -2)}{\nu -1}-1\\ & \notag -\frac{192 \beta
\kappa ^8 \nu  \phi ^4 \left(\frac{\phi }{M}\right)^{\nu -1}
\left(-\frac{12 \beta ^2 \kappa ^4 (\nu -1) \nu ^2
\left(\frac{\phi }{M}\right)^{2 \nu -3}}{M^3 \left(3 \gamma \kappa
^4+4 \beta  \kappa ^4 \left(\frac{\phi }{M}\right)^{\nu
}\right)^2}-\frac{3 \beta  (\nu -2) (\nu -1) \nu  \left(\frac{\phi
}{M}\right)^{\nu -3}}{M^3 \left(3 \gamma  \kappa ^4+4 \beta \kappa
^4 \left(\frac{\phi }{M}\right)^{\nu }\right)}\right)}{(\nu -1)^4
M \left(3 \gamma  \kappa ^4+4 \beta  \kappa ^4 \left(\frac{\phi
}{M}\right)^{\nu }\right)}
 \, ,
\end{align}
while the tensor spectral index reads,
\begin{align}\label{tensorspectralindexpowerlawmodelpower}
& n_{\mathcal{T}}\simeq \frac{2 \beta  \left(\nu  \left(2-3 \kappa
^2 \phi ^2\right)+2 \kappa ^2 \phi ^2+2 \nu ^3-4 \nu ^2\right)
\left(\frac{\phi }{M}\right)^{\nu }-3 \gamma  \kappa ^2 (\nu -1)
\phi ^2}{(\nu -1)^3 \left(3 \gamma +4 \beta  \left(\frac{\phi
}{M}\right)^{\nu }\right)}
 \, .
\end{align}
and the tensor-to-scalar ratio is,
\begin{equation}\label{tensortoscalarfinalmodelpowerlawpower}
r\simeq \frac{8 \kappa ^2 \phi ^2}{(\nu -1)^2}\, .
\end{equation}
The observational indices must be evaluated at the first horizon
crossing, so for $\phi=\phi_i$ and also the parameter $M$ is
assumed in this case too, to be of the form $M=\mu/\kappa$ where
$\mu$ is a dimensionless free parameter. The model at hand is
compatible with the Planck data for a wide range of the free
parameters. In Fig. \ref{plotplanck1power} we present the
marginalized curves of Planck 2018 and the predictions of Model
II, for the free parameters taking values $\mu=2.09\times
10^{-22}\times \kappa$, $\beta=[0.000025,1.69055]$,
$\gamma=10^{200}$, for $N=[50,63]$ $e$-foldings. As it can be seen
in Fig. \ref{plotplanck1}, the model is fitted quite well within
the Planck 2018 data. Also for this model too, the tensor spectrum
is blue tilted and the corresponding tensor spectral index takes
values in the range $n_{\mathcal{T}}=[0.00679,1.10833]$.
Let us now consider the evolution of Model II during the
post-inflationary era up to present day. For the numerical
analysis of the differential equations, we shall impose the
$y_H(z=z_{fin})=\frac{\Lambda}{3m_s^2}$,
$\frac{dy_H}{dz}\Big|_{z_{fin}}=0$, $\phi(z=z_{fin})=10^{-20}M_p$,
$\frac{d\phi}{dz}\Big|_{z=z_{fin}}=-10^{-10}M_p$, where
$z_{fin}=10^6$. The results of our numerical analysis for Model I
can be found in Fig. \ref{plot2power} where we plot the total
equation of state (EoS) parameter $w_{eff}$ (left plot) and the
deceleration parameter (right plot) as functions of the redshift
$z$.
As it can be seen in Fig. \ref{plot2} the total EoS parameter
approaches in this case too the value $w_{eff}\sim 1/3$, hence the
Model II for redshifts $z\sim 10^6$ has entered deeply in the
radiation domination era. Also the Model II mimics in this case
too the $\Lambda$CDM model at late times, by looking the behavior
of the deceleration parameter. In Table \ref{Table2} we have
gathered the values of the dark energy EoS parameter
$\omega_{DE}$, the dark energy density parameter $\Omega_{DE}$ and
of the deceleration parameter $q$ at present day. Overall, Model
II, as Model I, offers a viable description regarding its dark
energy predictions.
\begin{table}[h!]\caption{Model II Parameters}
\begin{center}
\begin{tabular}{|r|r|r|}
\hline \textbf{Parameter}&\textbf{Model II}&\textbf{$\Lambda$CDM Value or Planck 2018 Constraints}\\
\hline q(z=0)&-0.518953&-0.535
\\ \hline
 $\Omega_{DE}(0)$&0.679335&0.6847$\pm$0.0073\\ \hline
$\omega_{DE}(0)$&-1&-1.018$\pm$0.031\\ \hline
\end{tabular}
\end{center}
\label{Table2}
\end{table}
What now remains is to calculate the effect of the two
Einstein-Gauss-Bonnet models presented in this section, on the
energy spectrum of the primordial gravitational waves. This issue
is addressed in the next section.

\section{Primordial Gravitational Wave Energy Spectrum for Einstein-Gauss-Bonnet Theories}

In this section we shall calculate in detail the effect of the
GW170817-compatible Einstein-Gauss-Bonnet models we developed in
the previous sections on the energy spectrum of the primordial
gravitational wave energy spectrum. In the literature there exist
many works which consider theoretical predictions on primordial
gravitational waves, for a mainstream of articles see for example
Refs.
\cite{Boyle:2005se,Zhang:2005nw,Denissenya:2018mqs,Koh:2018qcy,Turner:1993vb,Schutz:2010xm,Sathyaprakash:2009xs,Caprini:2018mtu,
Kuroyanagi:2008ye,Clarke:2020bil,Nakayama:2009ce,Smith:2005mm,Giovannini:2008tm,
Liu:2015psa,Zhao:2013bba,Vagnozzi:2020gtf,Watanabe:2006qe,Kamionkowski:1993fg,Giare:2020vss,
Nishizawa:2017nef,Arai:2017hxj,Nunes:2018zot,Campeti:2020xwn,
Zhao:2006eb,Cheng:2021nyo,Chongchitnan:2006pe,Lasky:2015lej,Guzzetti:2016mkm,Capozziello:2017vdi,Odintsov:2021kup,Benetti:2021uea,Cai:2021uup,Lin:2021vwc,Zhang:2021vak,Odintsov:2021urx,Odintsov:2022cbm,Odintsov:2022sdk,Kawai:2021edk,Kawai:2021bye,Kawai:2017kqt,Moretti:2021ljj,Moretti:2022xem,Moretti:2020kpp}
and references therein. The results of our study can be compared
with the results of Ref. \cite{Koh:2018qcy} where also the
Einstein-Gauss-Bonnet primordial gravitational waves are
considered. The new development that our article brings in the
field is the detailed calculation of the effects that the
Einstein-Gauss-Bonnet term brings along to the energy spectrum of
the primordial gravitational waves, from present time back to the
radiation domination era. As we show shortly, the results are
quite interesting and surprisingly simple.
\begin{figure}[h!]
\centering
\includegraphics[width=40pc]{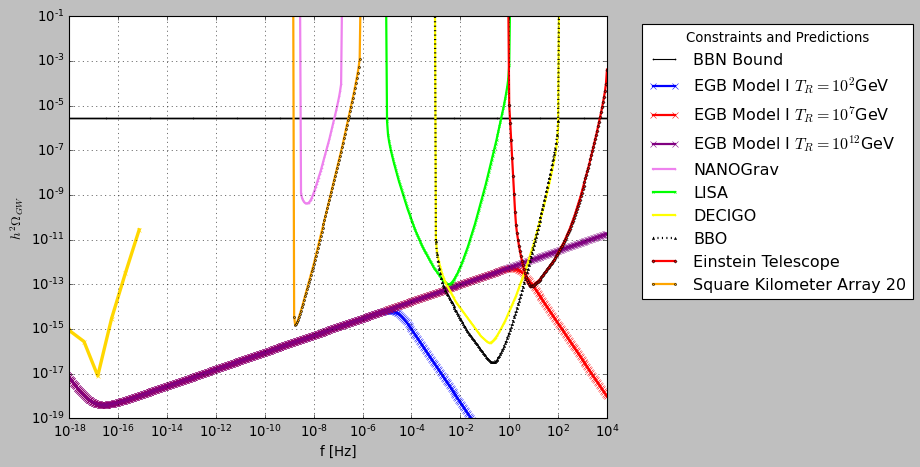}
\caption{The $h^2$-scaled gravitational wave energy spectrum for
the Einstein-Gauss-Bonnet Model I gravity. The
Einstein-Gauss-Bonnet Model I curves correspond to three different
reheating temperatures, the blue curve to $T_R=10^{2}$GeV, the red
curve to $T_R=10^{7}$GeV and the purple curve to
$T_R=10^{12}$GeV.} \label{plotfinalfrpure1}
\end{figure}
Let us present at this point how to quantify the effects of the
Einstein-Gauss-Bonnet gravity on the primordial gravitational
waveform. The evolution of the Fourier transformation of the
tensor perturbation of a flat FRW background has the following
form \cite{Odintsov:2021kup,Hwang:2005hb},
\begin{equation}\label{fouriertransformationoftensorperturbation}
\frac{1}{a^3Q_t}\frac{{\rm} d}{{\rm d} t}\left(a^3Q_t\dot{h}(k)
\right)+\frac{k^2}{a^2}h(k)=0\, ,
\end{equation}
where $Q_t$ for Einstein-Gauss-Bonnet gravity is defined below Eq.
(\ref{GW}), and we also quote its explicit form here for
convenience, $Q_t=F+\frac{Q_b}{2}$, $F=\frac{1}{\kappa^2}$ and
$Q_b=-8 \dot\xi H$. The evolution equation can be written as,
\begin{equation}\label{mainevolutiondiffeqnfrgravity}
\ddot{h}(k)+\left(3+\alpha_M
\right)H\dot{h}(k)+\frac{k^2}{a^2}h(k)=0\, ,
\end{equation}
where the parameter $\alpha_M$ is defined to be equal to,
\begin{equation}\label{amfrgravitypre}
a_M=\frac{Q_t}{Q_t H}\, .
\end{equation}
In the case of Einstein-Gauss-Bonnet gravity, the exact form of
the parameter $a_M$ can easily be calculated by using the explicit
form of $Q_t$,
\begin{equation}\label{amfrgravity}
a_M=\frac{-4\ddot{\xi}H-4\dot{\xi}H}{H(\frac{1}{\kappa^2}-4\dot{\xi}H)}\,
.
\end{equation}
Basically, the parameter $a_M$ quantifies the deviation of the
Einstein-Gauss-Bonnet primordial gravitational wave waveform from
that of standard GR. This can easily be measured by using a WKB
approach introduced in \cite{Nishizawa:2017nef,Arai:2017hxj}.
According to Refs. \cite{Nishizawa:2017nef,Arai:2017hxj}, by using
a WKB approach, the solution of the differential equation
(\ref{mainevolutiondiffeqnfrgravity}) can be written in the
following way,
\begin{equation}\label{mainsolutionwkb}
h=e^{-\mathcal{D}}h_{GR}\, ,
\end{equation}
where $h_{GR}$ denotes the GR waveform which corresponds to the
case $a_M=0$. The quantity $\mathcal{D}$ is defined as,
\begin{equation}\label{dform}
\mathcal{D}=\frac{1}{2}\int^{\tau}a_M\mathcal{H}{\rm
d}\tau_1=\frac{1}{2}\int_0^z\frac{a_M}{1+z'}{\rm d z'}\, .
\end{equation}
So in order to find the actual effect of the Einstein-Gauss-Bonnet
theory on the energy spectrum of the primordial gravitational
waves, one has to calculate the quantity $\mathcal{D}$ from
present day up to the reheating era, so for redshifts $z\sim
10^6$. This can easily be done by using the numerical outcomes of
the previous section regarding the behavior of the
Einstein-Gauss-Bonnet theory up to redshifts $z\sim 10^6$. In
addition to this, in the section where we calculated the
inflationary characteristics of the model, we also evaluated the
tensor spectral index and the tensor-to-scalar ratio, which is
also needed for the calculation of the energy spectrum of the
primordial gravitational wave. Let us recall in brief the form of
the energy spectrum of the primordial gravitational waves in the
context of GR, which is,
\begin{equation}
    \Omega_{\rm gw}(f)= \frac{k^2}{12H_0^2}\Delta_h^2(k),
    \label{GWspec}
\end{equation}
with $\Delta_h^2(k)$ being
\cite{Boyle:2005se,Nishizawa:2017nef,Arai:2017hxj,Nunes:2018zot,Liu:2015psa,Zhao:2013bba,Odintsov:2021kup},
\begin{equation}\label{mainfunctionforgravityenergyspectrum}
    \Delta_h^2(k)=\Delta_h^{({\rm p})}(k)^{2}
    \left ( \frac{\Omega_m}{\Omega_\Lambda} \right )^2
    \left ( \frac{g_*(T_{\rm in})}{g_{*0}} \right )
    \left ( \frac{g_{*s0}}{g_{*s}(T_{\rm in})} \right )^{4/3} \nonumber  \left (\overline{ \frac{3j_1(k\tau_0)}{k\tau_0} } \right )^2
    T_1^2\left ( x_{\rm eq} \right )
    T_2^2\left ( x_R \right ),
\end{equation}
and the quantity $\Delta_h^{({\rm p})}(k)^{2}$ denotes the
inflationary tensor power spectrum, which is,
\begin{equation}\label{primordialtensorpowerspectrum}
\Delta_h^{({\rm
p})}(k)^{2}=\mathcal{A}_T(k_{ref})\left(\frac{k}{k_{ref}}
\right)^{n_{\mathcal{T}}}\, .
\end{equation}
The inflationary tensor power spectrum has to be evaluated at the
Cosmic Microwave Background pivot scale
$k_{ref}=0.002$$\,$Mpc$^{-1}$ and also recall that
$n_{\mathcal{T}}$ stands for the spectral index  of the tensor
perturbations, while $\mathcal{A}_T(k_{ref})$ denotes the
amplitude of the tensor perturbations, the actual form of which
is,
\begin{equation}\label{amplitudeoftensorperturbations}
\mathcal{A}_T(k_{ref})=r\mathcal{P}_{\zeta}(k_{ref})\, .
\end{equation}
Finally $r$ is the tensor-to-scalar ratio and
$\mathcal{P}_{\zeta}(k_{ref})$ stands for the amplitude of the
primordial scalar perturbations. In effect we have,
\begin{equation}\label{primordialtensorspectrum}
\Delta_h^{({\rm
p})}(k)^{2}=r\mathcal{P}_{\zeta}(k_{ref})\left(\frac{k}{k_{ref}}
\right)^{n_{\mathcal{T}}}\, .,
\end{equation}
therefore the energy spectrum of the primordial gravitational
waves for Einstein-Gauss-Bonnet gravity takes the final form,
\begin{align}
\label{GWspecfR}
    &\Omega_{\rm gw}(f)=e^{-2\mathcal{D}}\times \frac{k^2}{12H_0^2}r\mathcal{P}_{\zeta}(k_{ref})\left(\frac{k}{k_{ref}}
\right)^{n_{\mathcal{T}}} \left ( \frac{\Omega_m}{\Omega_\Lambda}
\right )^2
    \left ( \frac{g_*(T_{\rm in})}{g_{*0}} \right )
    \left ( \frac{g_{*s0}}{g_{*s}(T_{\rm in})} \right )^{4/3} \nonumber  \left (\overline{ \frac{3j_1(k\tau_0)}{k\tau_0} } \right )^2
    T_1^2\left ( x_{\rm eq} \right )
    T_2^2\left ( x_R \right )\, ,
\end{align}
with the quantity $\mathcal{D}$ be defined in Eq. (\ref{dform}).
In effect, the impact of the Einstein-Gauss-Bonnet theory on the
GR energy spectrum of primordial gravitational waves is mainly
contained on the quantity $e^{-2\mathcal{D}}$ and also to the
inflationary observational indices $n_{\mathcal{T}}$ and $r$.
\begin{figure}[h!]
\centering
\includegraphics[width=40pc]{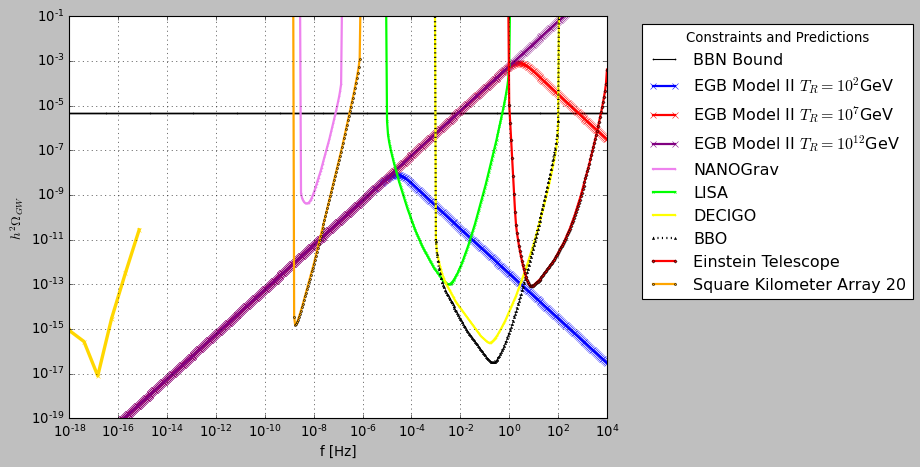}
\caption{The $h^2$-scaled gravitational wave energy spectrum for
the Einstein-Gauss-Bonnet Model II gravity. The
Einstein-Gauss-Bonnet Model II curves correspond to three
different reheating temperatures, the blue curve to
$T_R=10^{2}$GeV, the red curve to $T_R=10^{7}$GeV and the purple
curve to $T_R=10^{12}$GeV.} \label{plotfinalfrpure2}
\end{figure}
Now we proceed to the predictions of the energy power spectrum of
the primordial gravitational waves for the Einstein-Gauss-Bonnet
models I and II we developed in the previous sections. For these
models we calculated numerically the parameter $\mathcal{D}$ for
the redshift interval $z=[0,10^6]$ for both the models I and II.
The result is astonishing as it reveals a pattern, and for the
Model I we found that $\mathcal{D}=6.90102 \times 10^{-68}$ while
for the Model II we found that, $\mathcal{D}=6.9564\times
10^{-60}$. The remarkable result is that these are tiny for both
models, thus the amplification/damping factor is equal to unity
for both models I and II. Therefore, the overall amplification
effect of the Einstein-Gauss-Bonnet models which are compatible
with the GW170817 event is null, and this result of ours also
validates the approach developed in Ref. \cite{Koh:2018qcy}, where
the overall WKB factor $e^{-\mathcal{D}}$ was not taken into
account. As we showed, this is a correct approach since the
overall amplification is null. Now let us proceed in examining the
predictions of Model I and Model II regarding the energy spectrum
of the primordial gravitational waves. Our results are presented
in Fig. \ref{plotfinalfrpure1} for Model I while in Fig.
\ref{plotfinalfrpure2} we present the results for Model II,
regarding the $h^2$-scaled primordial gravitational wave energy
spectrum. In both models we considered three different reheating
temperatures, depicted in three distinct colors namely, the  blue
curve to $T_R=10^{2}$GeV, the red curve to $T_R=10^{7}$GeV and the
purple curve to $T_R=10^{12}$GeV. In all cases, for large
reheating temperatures, the signal from the Einstein-Gauss-Bonnet
models is detectable by the future experiments due to the fact
that for both models, the tensor spectral index is blue tilted and
significantly large. However, however for Model I, a low reheating
temperature will keep the gravitational wave signal undetectable.
Furthermore, the signal for Model II violates the BBN bound
constraints, for high frequencies, larger than $0.01$Hz. Thus the
Einstein-Gauss-Bonnet models, which are compatible with the
GW170817 event, serve as a potential viable candidate theory for
the description of the early Universe, if a stochastic
gravitational wave is detected by future experiments.

\section{Conclusions}

In this paper we calculated for the first time the overall
amplification/damping factor for the energy spectrum of the
primordial gravitational waves for Einstein-Gauss-Bonnet theories.
We focused on GW170817-compatible Einstein-Gauss-Bonnet theories,
and we presented the inflationary dynamics formalism for these
theories. In addition, we developed a formalism for the
post-inflationary evolution of GW170817-compatible
Einstein-Gauss-Bonnet theories, using the redshift as a dynamical
variable. In this way by numerically solving the field equations
using physically motivated initial conditions, we were able to
know in detail the dynamics of the Einstein-Gauss-Bonnet theories
at late times up to the reheating era. We applied the formalism to
two well chosen Einstein-Gauss-Bonnet models, and we demonstrated
that a viable inflationary era and a viable dark energy era can be
obtained by those models. In this way we provided a way toward a
unified description of early and late-time acceleration eras in
the context of Einstein-Gauss-Bonnet theories. Notably, both the
models we studied resulted to a blue tilted tensor spectral index,
and this has significant effects on the production of stochastic
primordial gravitational waves. Finally, we calculated the overall
effect of the Einstein-Gauss-Bonnet gravity on the
amplification/damping factor of the energy spectrum of the
primordial gravitational wave. As we showed, the amplification is
of the order of unity, and this validates the work of
\cite{Koh:2018qcy} who also considered similar
Einstein-Gauss-Bonnet theories, without the GW170817 constraint. A
natural extension of the present study is to include more quantum
corrections to the single scalar field Lagrangian, except from the
Gauss-Bonnet term only. For example one could consider power-law
curvature corrected Einstein-Gauss-Bonnet theories like in Ref.
\cite{Odintsov:2020ilr}. Work is in progress toward this research
line.

\end{document}